\documentclass[pre,aps,onecolumn,superscriptaddress,showpacs,floatfix]{revtex4}
\usepackage{epsfig,graphicx,latexsym}
\usepackage{amssymb,amsmath}

\begin{document}

\title{Approach to asymptotically diffusive behavior for Brownian particles in media with periodic diffusivities}
\author{David S. Dean}
\author{Thomas Gu\'erin}
\affiliation{Universit\'e de  Bordeaux and CNRS, Laboratoire Ondes et
Mati\`ere d'Aquitaine (LOMA), UMR 5798, F-33400 Talence, France}

\begin{abstract}
We analyze the mean squared displacement of a Brownian particle in a medium with a spatially varying local diffusivity which is assumed to be periodic. When the system is asymptotically diffusive the  mean squared displacement, characterizing the dispersion in the system, is, at late times, a linear function of time. A Kubo type formula is given for the mean squared displacement which allows the recovery of some known results for the effective diffusion constant $D_e$ in a direct way, but also allows an understanding of the asymptotic approach to the diffusive limit. In particular, as well as as computing the slope of a linear fit to the late time mean squared displacement, we find a formula for the constant where the fit intersects the $y$ axis.
\end{abstract}

\maketitle
\section{Introduction}

Many microscopic processes in solids and liquids can be described at a coarse grained level as  Langevin or Brownian processes \cite{satyabm}. The transport of Brownian particles subject to an external potential has been widely studied, the effect of the potential on the dispersion of the particle is to slow it down due to the trapping of the particle in local minima of the potential \cite{review,bouge}. Many transport processes are asymptotically diffusive, and the late time diffusion constant can be obtained by analyzing the mean squared displacement (MSD). Extracting information from particle trajectories is
a delicate business, and even the optimal way to estimate the diffusion constant for
pure Brownian motion is far from trivial \cite{fits}. However in systems where the diffusion is not a pure Brownian motion, in most cases the diffusive behavior only sets in at late times and short time transients make the statistical analysis of trajectories much more complicated and may even lead to behavior of the MSD similar to that seen for anomalous diffusion \cite{bouge}. Understanding finite time corrections to dispersive laws is thus potentially very important to correctly interpret experimental data.

In most studies, the local diffusion constant or diffusivity  is assumed to be constant meaning that the local mobility of the particle is assumed to be independent of its environment; only the local drift on the particle, generated by a spatially varying potential varies in space. For periodically varying bounded potentials, the mean squared displacement of the particle becomes linear at late times corresponding to normal diffusion. The late time diffusion constant, when it exists, is lower than the local diffusion constant, which turns out to be the short time diffusion constant, as diffusion dominates over drift at very short times due to the fact that it scales as $\sqrt{t}$, while drift scales as $t$. The effective diffusion constant for these models can be computed exactly in one dimension both without \cite{1d} and with an external bias \cite{tilt}. Recently, it was shown that the asymptotic approach to the final diffusive behavior could be studied in these systems via a Kubo-type formula for the MSD \cite{dea14}. In this paper we show how one can derive a Kubo formula for a Brownian particle with a spatially varying (but constant in time) diffusion coefficient (which is assumed to be periodic), in the case where there is no external potential acting on the particle. The formula we derive is general and valid in any dimension, however the quantities involved can, in general, only be exactly computed in one dimension.

The probability density function (pdf) of a  particle diffusing in a medium with a position dependent (but isotropic) diffusion constant (or diffusivity) $\kappa({\bf x})$  obeys the diffusion equation 
\begin{equation}
{\partial p({\bf x};t)\over \partial t} = \nabla\cdot\kappa({\bf x})\nabla p({\bf x};t).\label{fp}
\end{equation} 
This equation can be used to model a number of different situations where the local particle mobility, or equivalently diffusion constant, depends on the local environment.
An example is diffusion in porous media, where the effective diffusion constant is a function of the local pore size.  The most widely studied statistical quantity for such systems, which can be extracted from single particle trajectories, or from following particle ensembles, is the MSD. In systems where the asymptotic transport properties are normal (or classically diffusive) the MSD is given by
\begin{equation}
\langle ({\bf X}_t-{\bf X}_0)^2\rangle = 2dD(t) t, \label{defk}
\end{equation}
where  $D(t)$ is the effective time dependent diffusion constant and $d$ the spatial dimension. When the system  is 
asymptotically diffusive the effective, or late time, diffusion constant is defined via the limit
\begin{equation}
D_e = \lim_{t\to\infty}D(t).
\end{equation}
Clearly, for  pure Brownian motion where $\kappa({\bf x})=\kappa$ is independent of spatial position, the time dependent diffusion constant is constant in time and given simply by $D(t) = \kappa=D_e$.

\section{A Kubo-type formula for the MSD in a medium of varying diffusivity}\label{kubo}
In the Ito prescription of the stochastic calculus, the stochastic differential equation obeyed by particles described the the Fokker-Planck equation Eq. (\ref{fp}) is \cite{oks}
\begin{equation}
d{\bf X}_t = \sqrt{2\kappa({\bf X}_t)} d{\bf B}_t + \nabla \kappa({\bf X}_t) dt\label{sde}
\end{equation}
where $\langle dB_{it}dB_{jt}\rangle = \delta_{ij}dt$ and all increments $d{\bf B}_t$ at different times are independent and of mean zero. 
Integrating Eq. (\ref{sde}) yields
\begin{equation}
{\bf X}_t - {\bf X}_0 = \int_0^t ds\ \sqrt{2\kappa({\bf X}_s)} d{\bf B}_s + \int_0^t ds\  \nabla \kappa({\bf X}_s).
\end{equation}

The derivation of the Kubo formula is based on a simple trick, one subtracts the second term of the right hand side above from both sides and then squares the resulting equation to obtain:
\begin{equation}
 \langle ({\bf X}_t - {\bf X}_0)^2 - 2({\bf X}_t - {\bf X}_0)\cdot \int_0^t ds\  \nabla \kappa({\bf X}_s) + \int_0^t\int_0^t ds ds' \  \nabla \kappa({\bf X}_s)\cdot \nabla \kappa({\bf X}_{s'})\rangle
 =\langle \int_0^t\int_0^t ds ds'\ \sqrt{2\kappa({\bf X}_s)} \sqrt{2\kappa({\bf X}_{s'})}d{\bf B}_s\cdot d{\bf B}_{s'}\rangle. \label{int}
\end{equation}
We assume that the diffusion constant $\kappa({\bf x})$ is a periodic function in $d$ dimensions and we assume it is of period $l$ in all cartesian coordinates. Clearly, the methods presented here also apply to spaces divided into an infinitely repeating array of {\em elementary} unit cells $\Omega$ over which $\kappa$ is periodic, and which are not necessarily rectangular. We will now derive a formula that relates the MSD of a particle over all space to the properties of the diffusion in each individual cell $\Omega$. 

In order to proceed we assume that the system is of overall length $L\gg l$, where $L$ is an integer multiple of $l$ and that the system is periodic in $L$. We assume that ${\bf X}_0$ is uniformly distributed with pdf $p_0({\bf x}_0)= 1/L^d$ on the large periodic system. This choice of initial conditions will mean that the process ${\bf Y}_s= {\bf X}_s \ {\rm mod}\ l$ will be equilibrated and have initial conditions for ${\bf Y}_0$ given by the pdf $\rho_0({\bf y})= 1/l^d$.
The integrated stochastic differential equation Eq. (\ref{int}) is clearly not valid for a periodic system, however, if the asymptotic diffusion constant is given by $D_e$, so as long as $\langle ({
\bf X}_t - {\bf X}_0)^2\rangle \sim 2d D_e t \ll L^2$ the use of Eq. (\ref{int}) is valid.
Furthermore the time scale on which the asymptotic diffusive regime sets in will be determined by the length scale $l$ and given by $\tau \sim l^2/D_e$.
We now proceed to evaluate the various terms in Eq. (\ref{int}). By definition we have that
\begin{equation}
\langle ({\bf X}_t - {\bf X}_0)^2\rangle = 2dD(t) t.
\end{equation}
The second term on the left hand side of Eq. (\ref{int}) can be evaluated using 
\begin{eqnarray}
&&\langle ({\bf X}_t - {\bf X}_0)\cdot \int_0^t ds\  \nabla \kappa({\bf X}_s) \rangle \cr
&=& 
\int_0^t ds \left\{\ \int d{
\bf x}d{\bf y} d{
\bf x}_0 \ p({\bf x},{
\bf y}; t-s)p({\bf y},{\bf x_0};s) p_0({\bf x}_0){\bf x}\cdot\nabla\kappa({\bf y}) 
-\int d{\bf y}d{\bf x}_0 \ p({\bf y},{\bf x_0};s) p_0({\bf x}_0) {\bf x_0}\cdot\nabla\kappa({\bf y})\right\}
\label{term2}
\end{eqnarray}
where $p({\bf x},{\bf x}_0; s)$ is the transition probability density or propagator for a particle started at time $0$ at the point ${\bf x}_0$ to be at ${\bf x}$ at time $s$. Here it is useful to note that the propagator $p({\bf x},{\bf y};t)$ obeys the diffusion equation
\begin{equation} 
{\partial p\over \partial t}= - Hp\label{op}
\end{equation}
where $H$  is the self-adjoint operator defined by its action on  a function $f$ by
\begin{equation}
Hf({\bf x}) = -\int d{\bf y}\nabla \cdot \kappa({\bf x})\nabla \delta({\bf x}-{\bf y})f({\bf y})
\end{equation}
The initial condition for Eq. (\ref{op}) is given by $p({\bf x},{\bf y};0)=
\delta({\bf x}-{\bf y})$ and with this initial condition the propagator can be written in operator
notation as
\begin{equation}
p({\bf x},{\bf y}; t)=\exp(-tH)({\bf x},{\bf y}).
\end{equation}
As $H$ is self-adjoint, we also note that $\exp(-tH)$ is self adjoint and thus $p({\bf x},{\bf y}; t)=p({\bf y},{\bf x}; t)$. This symmetry of the transition density is crucial for the evaluation of the left hand side of Eq. (\ref{term2}). We first note that as $p_0({\bf x})$ is the equilibrium distribution for the process we have that, 
\begin{equation}
\int d{\bf x_0}p({\bf y},{\bf x_0};s) p_0({\bf x}_0)=p_0({\bf y}). 
\end{equation}
Using this property in Eq. (\ref{term2}), we obtain:
\begin{align}
\langle ({\bf X}_t - {\bf X}_0)&\cdot \int_0^t ds\  \nabla \kappa({\bf X}_s) \rangle \nonumber\\
&= \int_0^t ds\ \left\{\int d{
\bf x} d{
\bf y} \ p({\bf x},{
\bf y}; t-s)p_0({\bf y}){\bf x}\cdot\nabla\kappa({\bf y}) 
-\int d{\bf y}d{\bf x}_0 \ p({\bf y},{\bf x_0};s) p_0({\bf x}_0) {\bf x_0}\cdot\nabla\kappa({\bf y})\right\}
\label{term22}
\end{align}
Now, changing the time integration variable in the first term from $t-s$ to $s$ and relabeling the spatial integration variables and using the symmetry of the transition density we see that the term in Eq. (\ref{term22}) is identically zero. 

As for the third term on the left hand side of Eq. (\ref{int}), we use the fact that $\kappa({\bf X}_t)=\kappa({\bf Y}_t)$. The process ${\bf Y}_t$ has the same  Fokker-Planck  equation as ${\bf X}_t$ but where the operator $H$ acts on functions with ${\bf x} \in \Omega \equiv [0,l]^d$ with periodic boundary conditions. We thus find 
\begin{eqnarray}
\langle \int_0^t\int_0^t ds ds' \  \nabla \kappa({\bf X}_s)\cdot \nabla \kappa({\bf X}_{s'})\rangle 
&=& 2\int_0^t ds\int_0^s ds' \int_\Omega d{\bf x} d{\bf y}\  p({\bf x},{\bf y};s-s')p_0({\bf y})\nabla\kappa({\bf x})\cdot \nabla \kappa({\bf y})\nonumber \\
&=&2\int_0^t ds\int_0^s ds' \int_\Omega d{\bf x} d{\bf y}\  \exp\left(-(s-s')H\right)({\bf x},{\bf y}) p_0({\bf y})\nabla\kappa({\bf x})\cdot \nabla\kappa({\bf y}).
\end{eqnarray}
The operator $\exp(-tH)$ on $\Omega$ has an eigenfunction decomposition of the form
\begin{equation}
\exp(-tH)({\bf x},{\bf y}) = {1\over l^d} + \sum_{\lambda>0} \exp(-\lambda t)
\psi_\lambda({\bf x})\psi_\lambda({\bf y}).
\end{equation}
The first term above corresponds to the equilibrium distribution, and the following sum is 
over the non-zero eigenvalues $\lambda$ of $H$ with eigenfunctions $\psi_\lambda({\bf x})$. Due to the periodicity of $\kappa$, the term coming from the equilibrium distribution 
does not contribute and it is now straightforward to see that
\begin{equation}
\langle \int_0^t\int_0^t ds ds' \  \nabla \kappa({\bf X}_s)\cdot \nabla \kappa({\bf X}_{s'})\rangle 
= {2\over l^d}\int_\Omega d{\bf x}d{\bf y} \left[ \sum_\lambda \left({t\over \lambda} -{1-\exp(-\lambda t)\over \lambda^2}\right)\psi_\lambda({\bf x})\psi_\lambda({\bf y})\right]\nabla\kappa({\bf x})\cdot \nabla\kappa({\bf y}).
\end{equation}
The equivalent expression in operator notation is:
\begin{equation}
\langle \int_0^t\int_0^t ds ds' \  \nabla \kappa({\bf X}_s)\cdot \nabla \kappa({\bf X}_{s'})\rangle 
= {2\over l^d}\int_\Omega d{\bf x}d{\bf y} \left[ t H'^{-1}({\bf x},{\bf y}) - H'^{-2}({\bf x},{\bf y}) + H'^{-2}\exp(-t H')({\bf x},{\bf y})\right]
\nabla\kappa({\bf x})\cdot \nabla\kappa({\bf y}),
\end{equation}
where $H'$ denotes the operator $H$ acting on the subspace of functions orthogonal to the zero eigenvalue eigenfunction $\psi_0({\bf x}) = 1/\sqrt{l^d}$.  The operator $H'^{-1}$ is thus the pseudo-Green's function for $H$. 

To complete the computation we evaluate the right hand side of Eq. (\ref{int}). As we are using the Ito prescription of stochastic calculus, we have that for $s>s'$ that the increment $d{\bf B}_s$ is independent of $d{\bf B}_{s'}$, ${\bf X}_{s'}$ and also ${\bf X}_s$ -- consequently we have that
for $s\neq s'$
\begin{equation}
\langle \sqrt{2\kappa({\bf X}_s)} \sqrt{2\kappa({\bf X}_{s'})}d{\bf B}_s\cdot d{\bf B}_{s'}\rangle =0.
\end{equation}
However for $s=s'$ we have
\begin{equation}
\langle 2\kappa({\bf X}_{s})[d{\bf B}_s]^2\rangle =2d\langle \kappa({\bf X}_{s})\rangle\  ds
\end{equation}
where on the right hand side above the remaining average is with respect to the position ${\bf X}_s$. However as $\kappa({\bf X}_{s})=\kappa({\bf Y}_{s})$ and ${\bf Y}_s$ is in equilibrium with a uniform distribution we find that
\begin{equation}
\langle 2\kappa({\bf X}_{s})[d{\bf B}_s]^2\rangle =2d\overline\kappa\  ds,
\end{equation}
where $\overline \cdot$ indicates the spatial average of an $l$ periodic function over the region $\Omega$ and so
\begin{equation}
\overline\kappa = {1\over l^d}\int_\Omega d{\bf x}\ \kappa({\bf x}).
\end{equation}

Finally collecting all these results together we obtain
\begin{equation}
D(t) = \overline \kappa - {1\over l^d d}\int_\Omega d{\bf x}d{\bf y} H'^{-1}({\bf x},{\bf y})
\nabla\kappa({\bf x})\cdot \nabla\kappa({\bf y}) + {1\over d l^d t}\int_\Omega d{\bf x}d{\bf y} \left[ H'^{-2}({\bf x},{\bf y}) - H'^{-2}\exp(-t H')({\bf x},{\bf y})\right]
\nabla\kappa({\bf x})\cdot \nabla\kappa({\bf y})\label{kubo}.
\end{equation}
This formula relates explicitly the diffusion coefficient $D(t)$ in infinite space to the 
 properties of the diffusion on the unit domain $\Omega$.

Without going into explicit computations we can make a few general observations about the time dependent diffusion constant $D(t)$ from Eq. (\ref{kubo}). First, at short times
it is straight forward to see that
\begin{equation}
D(t) = \overline \kappa +\mathcal{O}(t),
\end{equation}
thus the short-time diffusion constant is given by the average diffusivity, this is easy to understand as at any time (given our assumption that ${\bf Y}$ starts in equilibrium) the distribution of the process ${\bf Y}_s$ is uniform. Recall, that in  the late time limit, we define the asymptotic diffusion constant as $D_e = \lim_{t\to\infty}D(t)$ and hence in Eq. (\ref{kubo}), only the first two terms contribute to this limit yielding
\begin{equation}
D_e = \overline \kappa - {1\over l^dd}\int_\Omega d{\bf x}d{\bf y} H'^{-1}({\bf x},{\bf y}) 
\nabla\kappa({\bf x})\cdot \nabla\kappa({\bf y}).\label{ke}
\end{equation}
We note that as $H'$ has positive eigenvalues we must have that $D_e < \overline \kappa$, and thus the late time diffusion is slower than one would naively expect. The formula of Eq. (\ref{ke}) agrees with two alternative derivations for $D_e$ in the literature. The first such derivation \cite{derr,review} was obtained by generalizing a method proposed by Derrida \cite{derro} for discrete random walks to the continuum case. Another method is based on the Kubo formula corresponding to the case for diffusion in a potential and a random time change argument that relates this problem to the varying spatial diffusion coefficient problem under consideration here \cite{review}. However, none of these approaches has to date been extended to examine finite time correction for the problem considered here.

The last term in Eq. (\ref{kubo}) represents the  finite time corrections to the asymptotic diffusion constant. However, the second term of this correction will decay exponentially quickly as
$\exp(-\lambda_0 t)$ where $\lambda_0$ is the lowest  non-zero eigenvalue of $H$. We expect  that $\lambda_0 \sim 1/l^2$ and thus for $t>l^2$, the time after which a particle  has typically  diffused through a few periods, we will find
\begin{equation}
D(t) = D_e + {C\over t},\label{asymp}
\end{equation}
with $C$ given by
\begin{equation}
C =  {1\over l^dd}\int_\Omega d{\bf x}d{\bf y} H'^{-2}({\bf x},{\bf y})\nabla\kappa({\bf x})\cdot \nabla\kappa({\bf y}).\label{ExpressionC}
\end{equation}
This means that the MSD behaves as 
\begin{equation}
\langle ({\bf X}_t-{\bf X}_0)^2\rangle  = 2d D_e t + 2d C.
\end{equation}
Thus the late time part the MSD can be fitted by a straight line, the slope of which is $2dD_e$ and the intercept with the $y$ axis is $2dC$.

Interestingly, the dependence of the operator $H$ on $\kappa$, means that, from Eqs. (\ref{ke},\ref{ExpressionC}),  multiplying $\kappa(x)$ by a constant leaves the constant $C$ unchanged, while the asymptotic diffusion constant $D_e$ is multiplied by the same constant. Thus, the constant $C$ is insensitive to the magnitude of the function $\kappa({\bf x})$, but depends only on its relative variations. 

%We also observe that if we write $\kappa({\bf x}) = \overline \kappa w({\bf x})$ where
%$\overline w =1$ we find that
%\begin{equation}
%D_e = \overline \kappa w_e 
%\end{equation}
%where $w_e$ is defined as the effective diffusion constant when $\overline \kappa=1$. However we find that
%\begin{equation}
%C = C_e,
%\end{equation}
%where $C_e$ the value of $C$ constant when $\overline \kappa=1$, that is to say that $C$ is independent of any constant rescaling of $\kappa({\bf x})$. 

\begin{figure}[t]
   \centering
  \includegraphics[width=12cm]{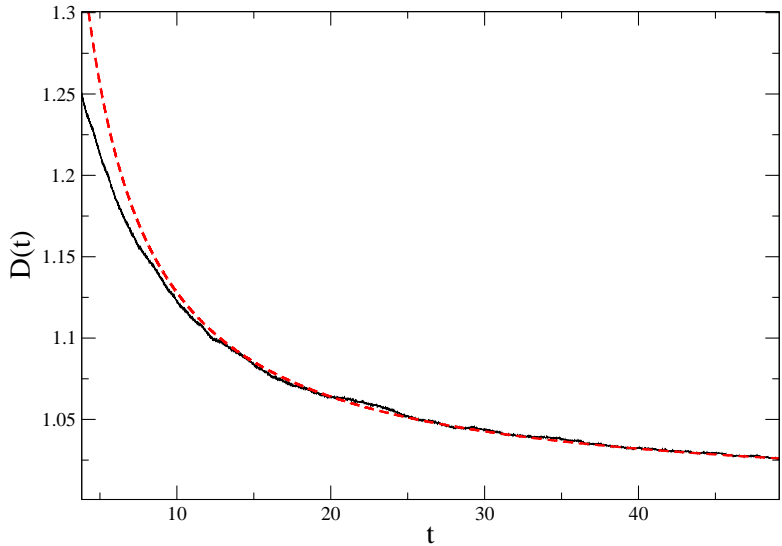}
  \caption{ (color online) $D(t)$ estimated from the MSD in a  numerical simulation of $5\times10^5$ Langevin particles, in one dimension, with local diffusion coefficient $\kappa(x)$ given by Eq. (\ref{kcos}) with $\alpha=0.8$ and $l=4\pi l_0$ (continuous) black line (shown in units such that $\kappa_0=1$ and $l_0=1$). Shown as thick dashed (red -color online) late time prediction of $D(t)$ of the form predicted by Eq. (\ref{asymp}) $D(t) =D_e + C/t$, where theory predicts that $D_e=1$ and $C=1.28$.   }
  \label{simul}
\end{figure}
The explicitly evaluation of $D_e$ and $C$ can only be carried out in one dimension and this will be done in the following section (\ref{1d}). However, in general we can write
the effective diffusion constant as
\begin{equation}
D_e = \overline \kappa - {1\over l^dd}\int_\Omega d{\bf x}
\nabla\kappa({\bf x})\cdot {\bf f}({\bf x}),\label{eqD}
\end{equation}
where ${\bf f}({\bf x})$ is given by
\begin{equation}
{\bf f}({\bf x})=\int_\Omega d{\bf y} H'^{-1}({\bf x},{\bf y}) \nabla\kappa({\bf y}),
\end{equation}
and is thus the solution to the equation
\begin{equation}
\nabla\cdot\kappa({\bf x})\nabla{\bf f}({\bf x}) = -\nabla\kappa({\bf x}). \label{EquationForf}
\end{equation}
The vector field ${\bf f}$ is periodic with period $l$ in all coordinates and in addition is
in the space orthogonal to $\psi_0({\bf x}) \propto 1$, which means that
\begin{equation}
\int_\Omega d{\bf x}\ {\bf f}({\bf x}) = {\bf 0}.
\end{equation}
The constant $C$ is then simply given by 
\begin{equation}
C = {1\over l^dd}\int_\Omega d{\bf x}\ {\bf f}^2({\bf x}). \label{LinkCAndF}
\end{equation}
Most results in the literature on effective diffusion constants are obtained via perturbative approaches 
\cite{dru87,dew95,abr95}. Interestingly, the functional renormalisation group \cite{dea08} can also be used to give approximate results which however agree with all known exact results for this problem. Exact results, where they exist, rely on the physical assumption that the self and collective diffusion constants are the same and thus the effective diffusion constant can be derived from the definition of $D_e$ given by
\begin{equation}
\overline {\bf j} = -D_e\overline{\nabla\rho}
\end{equation}
relating the spatially averaged current to the average particle gradient. The formula given here agrees with this definition \cite{review} and thus is a direct proof of this reasonable physical hypothesis. Using this definition for $D_e$ means that the one dimensional problem can be easily solved, while in certain cases a clever duality argument can give exact results in two dimensional systems \cite{dyk71,review}. In principle the formulas given here can be analyzed via perturbation expansions to give results for $D_e$ and $C$.

We emphasize that the relations given here are exact, however, in general,  the partial differential equation Eq. (\ref{EquationForf}) in two dimensions and 
higher is not amenable to analytic solution. However, in principle, Eq. (\ref{EquationForf}) should be easily amenable to numerical solution as the domain over which it holds is compact. The resulting integral expressions for $D_e$ (Eq. (\ref{eqD})) and $C$ (Eq. (\ref{LinkCAndF})) would then be straightforward to compute numerically.

\section{Exact results in one dimension}\label{1d}
In one dimension the function $f$ can be easily computed and one finds the general solution of Eq. (\ref{EquationForf}):
\begin{equation}
f(x) = -x + A + B\int_0^x {dy\over \kappa(y)} ,
\end{equation}
where $A$ and $B$ are constants. The condition of periodicity in $l$ then gives
\begin{equation}
B = \overline {\kappa^{-1}}^{-1} ,
\end{equation}
i.e. $B$ is the Harmonic mean of $\kappa$ under spatial averaging. The constant $A$ is then given by the condition $\int_0^l dx\  f(x) =0$ which gives 
\begin{equation}
A = {1\over l}\int_0^l dx\ \left[ x -  \overline {\kappa^{-1}}^{-1}\int_0^x dy \ \kappa^{-1}(y)\right].
\end{equation} 
If we define the function
\begin{equation}
r(x) = -x + \overline {\kappa^{-1}}^{-1}\int_0^x dy \ \kappa^{-1}(y),
\end{equation}
then the function $f(x)$ can be written as 
\begin{equation}
f(x) = r(x) - \overline r,
\end{equation}
where
\begin{equation}
\overline r = {1\over l}\int_0^l dx\ r(x) .
\end{equation}
Inserting this expression into Eq.  (\ref{LinkCAndF}), we obtain that the constant $C$ is equal to the variance of the function $r(x)$ with respect to spatial averaging:
\begin{equation}
C = {1\over l}\int_0^l dx \ (r(x)-\overline r)^2.
\end{equation} 
Using the solution for $f$  we find that the effective diffusion constant in one dimension is given by
\begin{equation}
D_e = \overline {\kappa^{-1}}^{-1},
\end{equation}
that is to say the Harmonic mean. This result for the one-dimensional diffusion constant is well known. The result is usually established by examining the steady state current in a system subjected to an applied external gradient, mathematically the diffusion constant is the same as the effective conductivity, effective dielectric permittivity or effective porosity of random media when $\kappa({\bf x})$ represents, respectively, a spatially varying conductivity, dielectric constant or porosity (appearing in Darcy's law for incompressible flow in porous media) \cite{review}.

If we make the periodicity of the diffusion constant explicit by writing 
\begin{equation}
\kappa(x) = K({2\pi x\over l})
\end{equation}
we find that the $l$ dependence of $C$ can be made explicit as 
\begin{equation}
C ={l^2\over (2\pi)^3}\int_0^{2\pi} dz \left(R(z)-\overline R\right)^2,
\end{equation}
where 
\begin{equation}
R(z) = -y +{1\over \overline {K^{-1}}}\int_0^y dy'\ K^{-1}(y'),
\end{equation}
where ${\overline K^{-1}}=\int_0^{2\pi} dy\ K^{-1}(y)/2\pi$ is again the inverse of the harmonic mean of the diffusion coefficient and ${\overline R}=\int_0^{2\pi} dy\ R(y)/2\pi$.
We thus see, that in common with the case of diffusion in a periodic potential \cite{dea14}, the constant $C$ is proportional to $l^2$. Indeed it is not hard to see, by dimensional analysis or rescaling of the diffusion equation, that this must be the case.  

As an concrete example, consider the system with local diffusion constant given by
\begin{equation}
\kappa(x) = {\kappa_0\over 1 + \alpha\cos({2\pi x\over l})},\label{kcos}
\end{equation}
where we must have $|\alpha|<1$.
Here we find that
\begin{equation}
D_e =\kappa_0 \ {\rm and }\ C = {\alpha^2\l^2\over 8\pi^2}.
\end{equation}
As a test of our theory we have numerically simulated an ensemble of $5\times10^5$ particles obeying a discretized form of Eq. (\ref{sde}) with $dt=0.0001$. The particles are started in the interval $[0,l]$, with $l=4\pi$, with a uniform distribution. The values $\kappa_0=1$ and  $\alpha=0.8$ are taken in this numerical example. The estimate of 
$D(t)$ is then made by estimating the ensemble average in  Eq. (\ref{defk}) by averaging over the ensemble of  particle trajectories. We see in Fig. (\ref{simul})  that at late times the asymptotic theory 
developed here predicts the correct behavior of $D(t)$, even for times where the  value of $D(t)$ is still significantly larger that its asymptotic late time limit $D_e$. 

\begin{figure}[t]
   \centering
  \includegraphics[width=12cm]{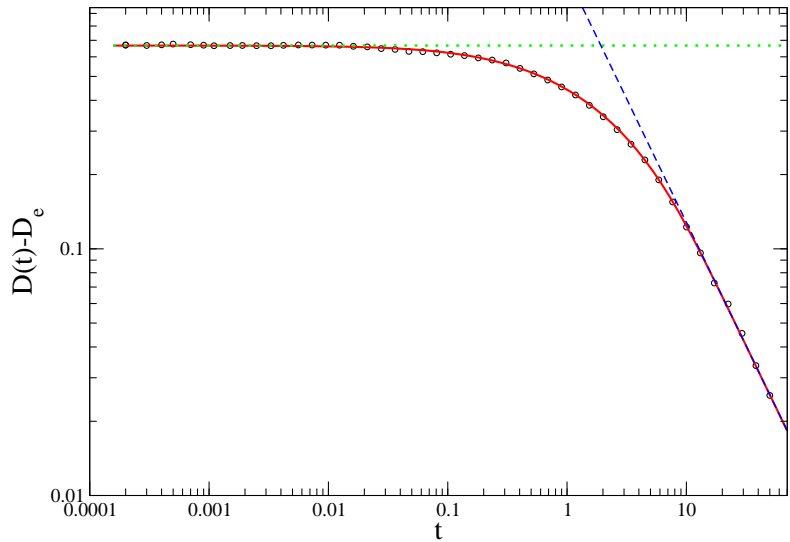}
  \caption{(color online) $D(t)-D_e$ (here $D_e=\kappa_0$) estimated from the MSD in the same  numerical simulation as Fig. (\ref{simul})  (open circles) compared at {\em all} times with a numerical evaluation of the Kubo formula Eq. (\ref{kubo}) (solid line- red online)(again shown in units such that $\kappa_0=1$ and $l_0=1$). Also shown is the asymptotic analytical late time correction $C/t$ to $D(t)-D_e$ (dashed line- blue online) as well as the short time value $D(0)-D_e = \overline\kappa-D_e$ (dotted line- green on line).}
  \label{num}
\end{figure}

The Kubo formula Eq (\ref{kubo}) in principle gives the behavior of the MSD at all times
assuming the initial distribution is the equilibrium one. Even in one dimension the full time 
dependent problem is not analytically tractable. However numerically one can construct 
a discrete approximation to the operator $H$ on a lattice subspace of $[0,l]$. The resulting operator can then be numerically analyzed to obtain the eigenfunctions $\psi_\lambda$ and corresponding eigenvalues $\lambda$. The last term in Eq. (\ref{kubo}) containing the exponentially suppressed terms can then be evaluated numerically. We have carried out this numerical computation for the system we simulated to obtain Fig. (\ref{simul}) and the comparison is shown in Fig (\ref{num}). We see that the full Kubo formula agrees perfectly with the results of the numerical simulation.

\section{Conclusions}
We have analyzed the temporal behavior of the MSD of Brownian particles diffusing in a medium with a spatially varying local diffusion constant. The assumption behind the calculation is that at the starting time the system is in equilibrium, with a uniform distribution, and we assume that the time at which the diffusion constant attains its final asymptotic limit is always smaller than the time for the particle to  {\em feel} the finite size of the system. Clearly by taking large enough systems this limit can always be achieved. At short times, as the position distribution is uniform, the diffusion constant is clearly given by
$\overline \kappa$. At late times the effective diffusion constant is smaller, $D_e < \overline\kappa$. This result is not as easy to understand intuitively as the slowing down of diffusion in an external potential, where trapping in the potential's minima clearly provides 
the mechanism by which dispersion is slowed down.   For a periodic local diffusivity, we have show that the late time correction to the time dependent diffusion constant, as defined by Eq. (\ref{defk}) decays as $C/t$ and we have given a general expression for the coefficient $C$. In one dimension, the effective diffusion constant only depends on the composition of the 
local diffusivity via its harmonic mean. The constant $C$ depends however on the spatial distribution of $\kappa$, thus two systems with the same value of $D_e=1/\overline{ \kappa^{-1}}$ will in general have differing values of $C$. The explicit analytical  results given here are restricted to the case of one dimension. However, in two dimensional random systems if we write $\kappa({\bf x}) =\kappa_0\exp\left(\phi({\bf x})\right)$ and if the corresponding field $\phi$ is statistically identical to $-\phi$ then the effective diffusion constant is given by $D_e=\kappa_0$ \cite{dyk71,review}. This result is derived by a duality argument in two dimensions and it would be interesting to see if this argument could be extended to compute $C$. 

In principle the method here could also be extended to analyze more general diffusion equations, for example one could extend the study here, and in Ref. \cite{dea14}, by considering diffusion in a periodic potential with a periodic diffusion constant in such a way that the underlying equilibrium distribution is the Gibbs-Boltzmann one (as detailed balance is obeyed). Finally, in this work, and the study of diffusion in a periodic potential \cite{dea14}, the steady state has an equilibrium distribution without a current. It would be very interesting to examine the temporal dependence of the MSD in systems having a steady state with current, for instance Brownian particles in a tilted potential \cite{tilt}.

\end{document}